\documentclass[reprint,
 amsmath,amssymb,
 aps,
]{revtex4-1}

\usepackage{graphicx}
\usepackage{dcolumn}
\usepackage{bm}
\usepackage{tensor} 

\begin{document}

\title{Is the Hubble diagram of quasars in tension with concordance cosmology? }

\author{Hermano Velten}%
 \email{hermano.velten@ufop.edu.br}
\affiliation{Departamento de F\'isica, Universidade Federal de Ouro Preto (UFOP), Campus Morro do Cruzeiro, 35400-000, Ouro Preto-MG Brazil}

\author{Syrios Gomes}%
 \email{syriosgs@gmail.com}
\affiliation{N\'ucleo Cosmo-UFES \& Departamento de F\'isica, CCE, Universidade Federal do Esp\'irito Santo, 29075-910, Vit\'oria-ES, Brazil}

\date{\today}
\begin{abstract}
\noindent
Recently Risaliti \& Lusso [Nature Astron. 3 (2019) 3 272] reported new measurements of the expansion rate of the Universe by constructing the Hubble diagram of 1598 quasars in the redshift range $0.5<z<5.5$. It is claimed a $4\sigma$ tension with the standard concordance $\Lambda$CDM concerning both the fractionary matter density $\Omega_{m0}$ and the dark energy equation of state parameter $w_{de}$ standard values. In this work we promote an independent analysis of the same data set using a model-independent estimator for cosmic acceleration. Our results corroborate that the source of such tension can be related to the $\Omega_{m0}$ value with a reasonable indication of a higher $\Omega_{m0}$ value ($\Omega_{m0} \gtrsim 0.4$). On the other hand, we find that the role played by $w_{de}$ on the claimed tension is weak. We also discuss the use of this estimator as a ``quality tool'' to test the robustness of Hubble diagrams. We conclude claiming that the Quasars data can not yet be seen  as a reliable cosmological tool since they can not even state the universe experienced an accelerated expansion phase. 
\begin{description}
\item[PACS numbers] 98.80.-k, 98.54.Aj
\end{description}
\end{abstract}

\maketitle

\section{Introduction}

The necessary ingredients to build the late time background evolution of the standard flat cosmological model are the local (or today's) Hubble constant $H_0$ and the amount of pressureless matter, given by the fractionary density parameter $\Omega_{m0}=\rho_m/\rho_{c0}$, where $\rho_{c0}=3H^{2}_0/8\pi G$ is the today's critical density. Another possible degree of freedom is equation of state of dark energy $w_{de}=P_{de}/\rho_{de}$, where $P_{de}$ ($\rho_{de}$) is the dark energy pressure (energy density). This corresponds to the expansion rate $H=\dot{a}/{a}$, where the symbol $``^{.}"$ means derivative with respect to the cosmic time, of the type
\begin{equation}
H = H_0 \left[\frac{\Omega_{m0}}{a^{3}}+(1-\Omega_{m0})a^{-3(1+w_{de})}\right]^{1/2}.
\label{Eq1}
\end{equation} 
 The function $a(t)$ is the cosmic scale factor. Recent observations including Cosmic Microwave Background (CMB), Baryon acoustic oscillations and Supernovae (SN) type Ia data constrain this scenario close to the preferred parameters values $\Omega_{m0} \simeq 0.3$ and $w_{de}\simeq -1$. The latter result makes (\ref{Eq1}) equivalent to the so called flat $\Lambda$CDM model \cite{Aghanim:2018eyx}. 

Since the discovery that the width of the lightcurve peak of Supernovae type Ia (SNIa) is correlated with their peak luminosity (the so called Phillips calibration \cite{Phillips:1993ng}) such objects have been considered reliable standard candles and widely used in observational cosmology. Hence, the SN type Ia Hubble diagram (SNHD) represents one of the observational pillars supporting the cosmological concordance model. However, in terms of the redshift range distribution, even recent SNHD catalogues are limited to redshifts up to $z\sim 2$. Having SN data in the range $1<z<2$ make such samples deep enough to state cosmic acceleration but somehow limited to probe the early background evolution. Then, the search for alternative reliable candles in redshift range beyond the SNHD ($z > 2$) opens the possibility to test cosmology dynamics deep in the universe lifetime at moments between the early Universe and the last scattering surface $z\sim 1100$ where CMB has been released. There are phenomenological attempts in the literature to build a Hubble diagram for objects that are not considered standard candles but are at such high redshifts. This includes Gamma Ray Bursts (GRBs) \cite{Demianski:2016zxi,Wei:2013xx,Izzo:2009bw,Cardone:2009mr,Liang:2008kx,Schaefer:2006pa} and Quasars \cite{Risaliti:2018reu,Lusso:2017hgz, Risaliti:2016nqt, Risaliti:2015zla}. Mostly of these approaches are based on the strategy to find out empirical correlations between observational properties of these objects e.g., fluxes at different wavelengths, emission peaks among others observational quantities in a certain redshift range and then to extrapolate the fitting parameters to the entire sample. 

Concerning the Quasars Hubble Diagram (QSOsHD), recently Risaliti \& Lusso presented a robust catalogue containing 1598 inferred luminosity distances of quasars in the redshift range $0.5<z<5.1$ \cite{Risaliti:2018reu}. The final catalogue represents the onset of a gradual refinement on the selection techniques and flux measurements developed along Refs. 
\cite{Lusso:2017hgz, Risaliti:2016nqt, Risaliti:2015zla}. The strategy performed in these works is based on a correlation between the X-ray and the optical-ultraviolet (UV) emissions at low redshift i.e., there exists a phenomenological parametrization such that $log (L_X) = \gamma log (L_{UV})+\beta$ where $\gamma$ and $\beta$ are fitting parameters of the luminosities at $2$keV ($L_{X}$) and $2500 \AA$ ($L_{UV}$) \cite{Avni}. Another recent work from this same group has shown that this relation persists to even higher redshifts up to $z\sim 7$ \cite{Salvestrini:2019thn} (see also \cite{Lusso:2019akb}). This corroborates therefore the strategy of Ref.\cite{Risaliti:2018reu}. The latter reference reports that while the low-z subsample (containing objects with redshifts $z<2.1$) of the QSOsHD is consistent with the standard cosmological model, the inclusion of the high-z subsample yields to a $\sim 4 \sigma$ tension with standard cosmology. By fitting the expansion (\ref{Eq1}) to the QSOsHD data it is found a preference for a higher fractionary matter density parameter $\Omega_{m0}>0.4$ and a phantomic dark energy $w_{de}<-1$. It is worth of mentioning that even some SN analysis have reported tensions with relation to the standard cosmology either by stating a marginal evidence for acceleration \cite{Nielsen:2015pga} or inferring a higher $\Omega_{m0}$ values \cite{Shariff:2015yoa}.

Our goal in this work is to detail the analysis of the QSOsHD by searching for the source of such tension in an independent way. Rather than data fitting and performing a parameter estimation via statistical techniques we employ an model-independent estimator for cosmic acceleration proposed in Refs. \cite{Seikel:2007pk,Seikel:2008ms}. This will allow us to answer two basic questions: Is there a tension between the QSOsHD data and the standard cosmological model? Is the QSOsHD a reliable observational source? 

In the next section we introduce the method and apply it to the data. We also use SNHD from Pantheon \cite{Scolnic:2017caz} sample as alternative data set of our analysis. This allows us to update the evidence for acceleration available in the SNHD and to promote a proper comparison with the QSOsHD. Our main results are presented in section III and we conclude in the final section.

\section{Evidence for acceleration in Hubble diagram}\label{sect2}

Our analysis is based on the estimator proposed in \cite{Seikel:2007pk, Seikel:2008ms} and revisited recently in Ref. \cite{Velten:2017ire}. Its main goal is to provide a quantitative measurement in terms of a statistical Confidence Level (CL) about the accelerated dynamics of the universe in a model independent way. The construction of such estimator is based on the definition of the deceleration parameter
\begin{equation}\label{qz}
q(z)= \frac{H^{\prime}(z)}{H(z)} (1+z) - 1,
\end{equation}
where a prime $`` \, ^{\prime} \,"$ denotes a derivative with respect to the redshift. An accelerated background expansion at a certain redshift is indicated by $q(z)<0$. From (\ref{qz}) one can obtain 
\begin{equation}
{\rm ln}\,\frac{H(z)}{H_0}=\int^z_0 \frac{1+q(\tilde{z})}{1+\tilde{z}} d\tilde{z}.
\label{lnH}
\end{equation}

The null hypothesis proposed in \cite{Seikel:2007pk} is that the universe has never expanded in a accelerated way. This means that $q(z)>0 \, \forall \, z$. See also Refs. \cite{Santos:2007pp,Gong:2007zf} for similar approaches.

By applying the latter inequality to Eq (\ref{lnH}) it turns into
\begin{equation}
{\rm ln}\,\frac{H(z)}{H_0}\geq\int^z_0\frac{1}{1+\tilde{z}}d\tilde{z}={\rm ln}\,(1+z).
\label{ineqLnH}\end{equation}
From the Hubble diagram one obtains the luminosity distance which for a homogeneous, isotropic and expanding background (a Friedman-Lemaitre-Robertson-Walker universe) reads
\begin{equation}
d_L(z)= (1+z) \int^{z}_0 \frac{d\tilde{z}}{H(\tilde{z})}.
\end{equation}
It is possible to recast inequality (\ref{ineqLnH}) via the definition $d_L$ such that it becomes
\begin{equation}
d_L\leq (1+z) \frac{1}{H_0} \int^z_0 \frac{d\tilde{z}}{1+\tilde{z}}=(1+z)\frac{1}{H_0} {\rm ln}(1+z). 
\end{equation}
The luminosity distance is essential to compute the observed distance modulus $\mu$, the quantity directly related to the observations,
\begin{equation}
\mu = m-M= 5 log (d_L / Mpc) + 25,
\label{mu}
\end{equation}
where $M$ and $m$ are the absolute and apparent magnitudes respectively.

For each object $i$ at a redshift $z_i$ in the Hubble diagram we can define the quantity
\begin{eqnarray}
&&\Delta \mu_{obs} (z_i) = \mu_{obs}(z_i) - \mu(q=0) \\ \nonumber
&=& \mu_{obs} (z_i)-5{\rm log}\left[\frac{1}{H_0}(1+z_i){\rm ln} (1+z_i)\right]-25,
\label{eqDeltaobs}
\end{eqnarray}
which is the difference between its observed distance modulus $\mu_{obs}^i$ and the distance modulus of a universe with constant deceleration parameter $q=0$ from today until the redshift $z_i$. Then, the estimator has been designed in (8) such that up to this point we have assumed only that light propagates on null geodesics in a homogeneous, isotropic and spatially flat universe.
Clearly, positive $\Delta \mu_{\rm th}$ values indicate acceleration. The face value of $\Delta \mu_{obs}$ is meaningless if its inferred error is not included. Taking into the account the error of the redshift ($\sigma_z$) and peculiar velocities ($\sigma_{v}$) of each SN, the error of the quantity $\Delta \mu_{obs}$ becomes

\begin{equation}
\sigma_i=\left[\sigma^2_{\mu_i}+\left(\frac{5\, ln\left(1+z_i\right)+1}{(1+z_i)\, ln (1+z_i) ln10}\right)^2\left(\sigma^2_z+\sigma^2_v\right)\right]^{1/2}.
\end{equation}

The so called ``single SN analysis" corresponds to computing the quantity $\Delta \mu_{obs}$ for each SN individually. An interesting output of this analysis is that even a few SN indicate that the universe never accelerated (i.e., $\Delta \mu_{obs} <0 $). However, as expected from the statistical nature of such analysis the majority of SN in usual Supernovae samples indicates acceleration. 

As already noticed by Refs. \cite{Seikel:2007pk, Seikel:2008ms} the single SN analysis if of limited statistical interest. A more interesting analysis of the estimator $\Delta \mu_{obs}$ is obtained with the so called ``averaged SN analysis". In this analysis we group a number $N$ of objects defining the mean value
\begin{equation}
\overline{\Delta\mu}=\frac{\sum_{i=1}^{N} g_i \,\Delta\mu_{obs}(z_i)}{\sum_{i=1}^{N} g_i},
\label{10}\end{equation}
where the factor $g_i = 1/ \sigma^{2}_i$ makes the data points with smaller errors contribute more to the average. The standard deviation of the mean value is defined by
\begin{equation}
\sigma_{\overline{\Delta\mu}}=\left[\frac{\sum^N_{i=1} g_i \left[\Delta \mu_{obs} \left(z_i\right) - \overline{\Delta\mu}\right]^2}{(N-1)\sum^N_{i=1}g_i}\right]^{1/2}.
\label{11}\end{equation}

The grouping criteria can obey either a fixed redshift range or a fixed object number $N$ per bin. The quantitative value for the evidence of acceleration in each SN bin is given by $\overline{\Delta\mu}$ divided by the error $\sigma_{\overline{\Delta\mu}}$.

One may now wonder what is the expected result for the estimator $\Delta \mu$ for a giving cosmology. In Ref. \cite{Velten:2017ire} we addressed this issue by simulating mock data for certain cosmologies and for specific Hubble diagram distributions. It has been verified that the estimator satisfactorily computes the evidence for acceleration in actual SNHDs like the JLA sample in comparison with simulated catalogues. The full JLA sample provides a $20.40 \sigma$ statistical CL asserting accelerated expansion (see first row of Table \ref{TableI}). This result appeared firstly in Ref. \cite{Velten:2017ire}. Columns in this table represent, from left to right, the sample studied, the evidence for acceleration, the number of objects in the sample and the mean redshift. 

We calculate now the evidence for acceleration in the PANTHEON sample. The construction of the HD with the Pantheon SN data \cite{Scolnic:2017caz} employs a different procedure in which the $H_0$ value is degenerated with the absolute magnitude. Then, the PANTHEON sample does not allow to constrain $H_0$. It is also necessary to set a fiducial cosmology to generate the Hubble diagram. This is why one can infer the evidence for expansion for different cosmologies.  We present these results in Table \ref{TableI}. It shows that the PANTHEON sample also provides strong evidence favoring acceleration for both the $\Lambda CDM$ model ($+23.73 \sigma$), a $w$CDM model ($+24.23 \sigma$) and a CPL dark energy parameterization ($+23.62 \sigma$). It is also worth noting that when substracting low-z data from such samples i.e., data at $z<0.1$ (as in Ref.\cite{Seikel:2008ms}), the evidence increases.

\begin{table}
\centering
\caption{Averaged evidence (in $\sigma$ of C.L.)  for acceleration for different SNHD samples. For the samples indicated with {\it no low-z} data at $z<0.1$ have been not considered.}
{\begin{tabular} {c||c||c||c}
Sample & $\overline{\Delta \mu}$ / $\sigma_{\overline{\Delta\mu}}$ & \# of & Mean \\
 & Evidence & Objects & redshift \\
\hline \hline 
FULL JLA &  +20.40  & 740 & 0.32 \\
JLA no low-z &  +22.33  & 588 & 0.40 \\
Pantheon $\Lambda$CDM&  +23.73 & 1048 & 0.32 \\
Pantheon $\Lambda$CDM no low-z&  +28.25 & 837 & 0.39 \\
Pantheon $w$CDM&  +24.23 & 1048 & 0.32 \\
Pantheon $w$CDM no low-z&  +28.73 & 837 & 0.39 \\
Pantheon CPL&  +23.62 & 1048 & 0.32 \\
Pantheon CPL no low-z&  +28.14 & 837 & 0.39 \\
\hline
\end{tabular}} 
\label{TableI}
\end{table}

\section{Results for the Hubble diagram of Quasars }

We apply the estimator described in the last section to the QSOsHD developed in Ref.  \cite{Risaliti:2018reu}. By using Eqs. (\ref{10}) and (\ref{11}) we obtain the results shown in Table \ref{TableII}. The result for the full QSOsHD sample containing 1598 objects reveals a surprisingly strong lack of evidence favoring acceleration ($-13.60 \sigma$). A similar results is obtained even substracting the low-z sample ($-13.55\sigma$). Indeed, this does not affect QSOs since there are only a few of them at $z<0.1$. We also split the QSOsHD into a subsample with objects at $z<1.3$, the same redshift range of the JLA SN sample. This sub-sample is also not able to provide evidence for acceleration ($-7.95 \sigma$). This splitting is motivated by the procedure adopted in Ref. \cite{Risaliti:2018reu}. The latter assumes a log-linear relation between the rest-frame monochromatic luminosities at 2KeV ($L_X$) and 2500 \AA in this redshift range. The parameters of this relation are found by a joint fit with the JLA SNHD sample. Hence, it can be said that the JLA sample is used to calibrate the QSOsHD at low redshifts. The obtained fitting parameters for the $log (L_X)  $ x $ log (L_{UV})$ relation are assumed to be valid to the entire QSOs data sample at higher redshifts. It has been also recently shown that there is no significant evolution of this correlation (fitting parameters) towards high redshifts \cite{Salvestrini:2019thn}. Therefore the method seems robust and pertinent for building a reliable HD of high-z objects. A similar technique has also been used to build the Gamma-Ray Bursts Hubble diagram \cite{Liang:2008kx}.

\begin{table}
\centering\caption{Averaged evidence (in $\sigma$ of C.L.) for acceleration for the QSOsHD. All results adopt $H_0 = 70.0$ km/s/Mpc. }
{\begin{tabular} {c||c||c||c}
Sample & $\overline{\Delta \mu}$ / $\sigma_{\overline{\Delta\mu}}$ & \# of  & Mean\\
 & Evidence& Objects & redshift \\
\hline \hline 
FULL Quasars &  -13.60 & 1598 & 1.34  \\
Quasars no low-z &  -13.55 & 1587 & 1.35  \\
QSOs $0<z\leq1.3$&  -7.95 & 968 & 0.80 \\
QSOs $1.3<z\leq5.1$&  -16.21 & 630 & 2.15\\
 \hline
\end{tabular}} 
\label{TableII}
\end{table}

Results shown in Table \ref{TableII} for the QSOsHD do not support standard cosmology. Rather than inferring deceleration, negative $\overline{\Delta \mu}/ \sigma_{\overline{\Delta \mu}}$ values mean that there is no indication for acceleration. This should be the proper interpretation of such results. Negative $\overline{\Delta \mu}/ \sigma_{\overline{\Delta \mu}}$ are also found even for the low-z subsample though Ref. \cite{Risaliti:2018reu} states that this sub-sample is in agreement with all the main current cosmological probes. This tension is actually caused by the large dispersion of the QSOs at low redshifts and due to the fact some ``far from the $\Lambda$CDM best fit'' data points have very small error bars contributing then more to the estimator.

\begin{figure}[t]
\includegraphics[width=0.48\textwidth]{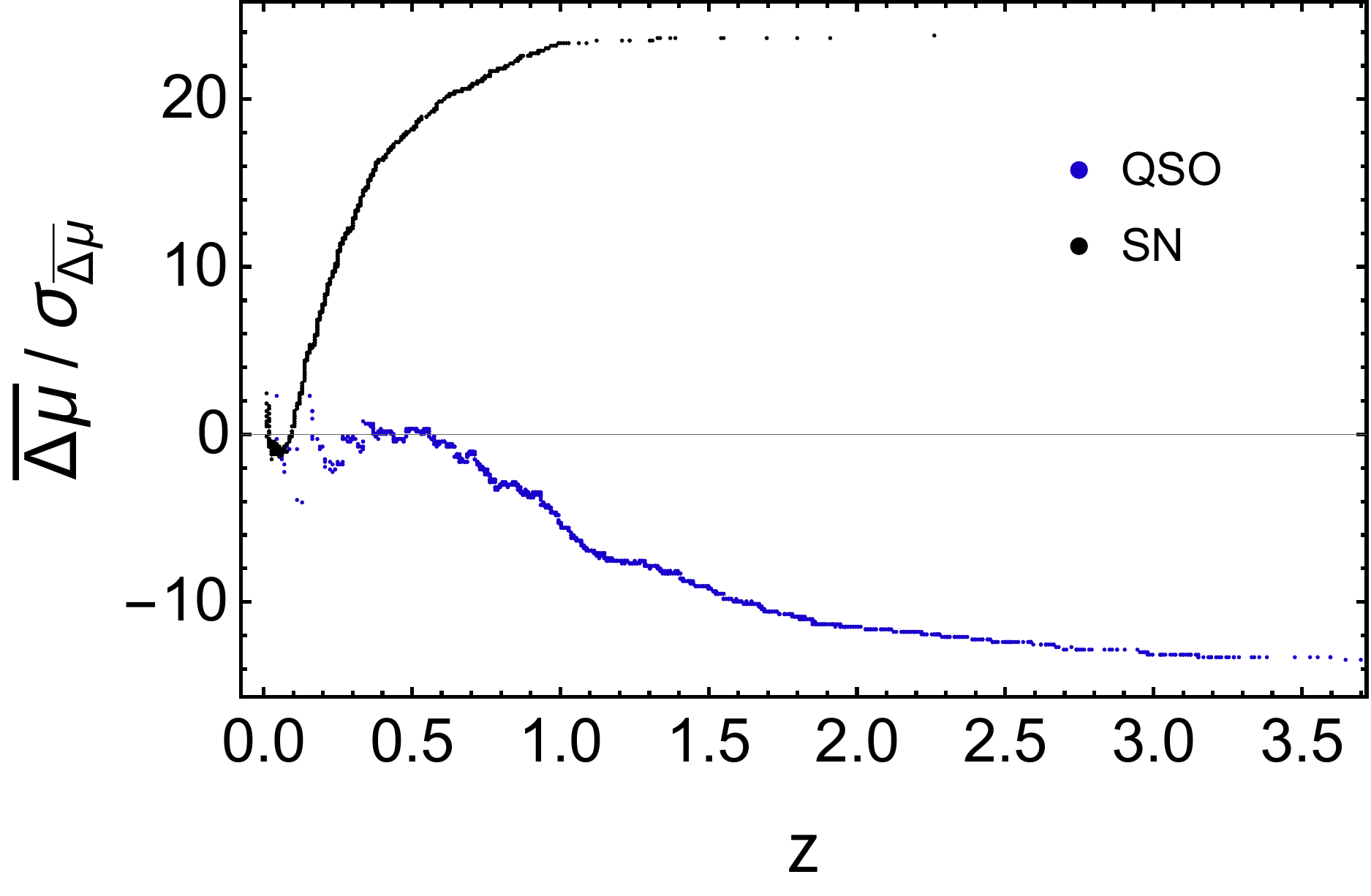}
\caption{Cumulative evidence for the SN PANTHEON (black) sample and QSOsHD (blue).     }
\label{FigCumulative}
\end{figure}

 Fig. \ref{FigCumulative} shows how the cumulative $\Delta \mu$ evolves with the redshift. Black (Blue) dots follow the PANTHEON (QSOs) sample. For the SN data this analysis reveals that indeed high redshift information is demanded in order to establish positive evidence for acceleration. On the other hand, the cumulative $\Delta \mu$ for the QSOsHD clearly evolves to negative values as higher redshifts are taken into account. This confirms that this sample is not able to infer acceleration.

We investigate now on possible sources for such discrepancy i.e., why does not the QSOsHD provide evidence for acceleration?

\begin{figure}[t]
\includegraphics[width=0.48\textwidth]{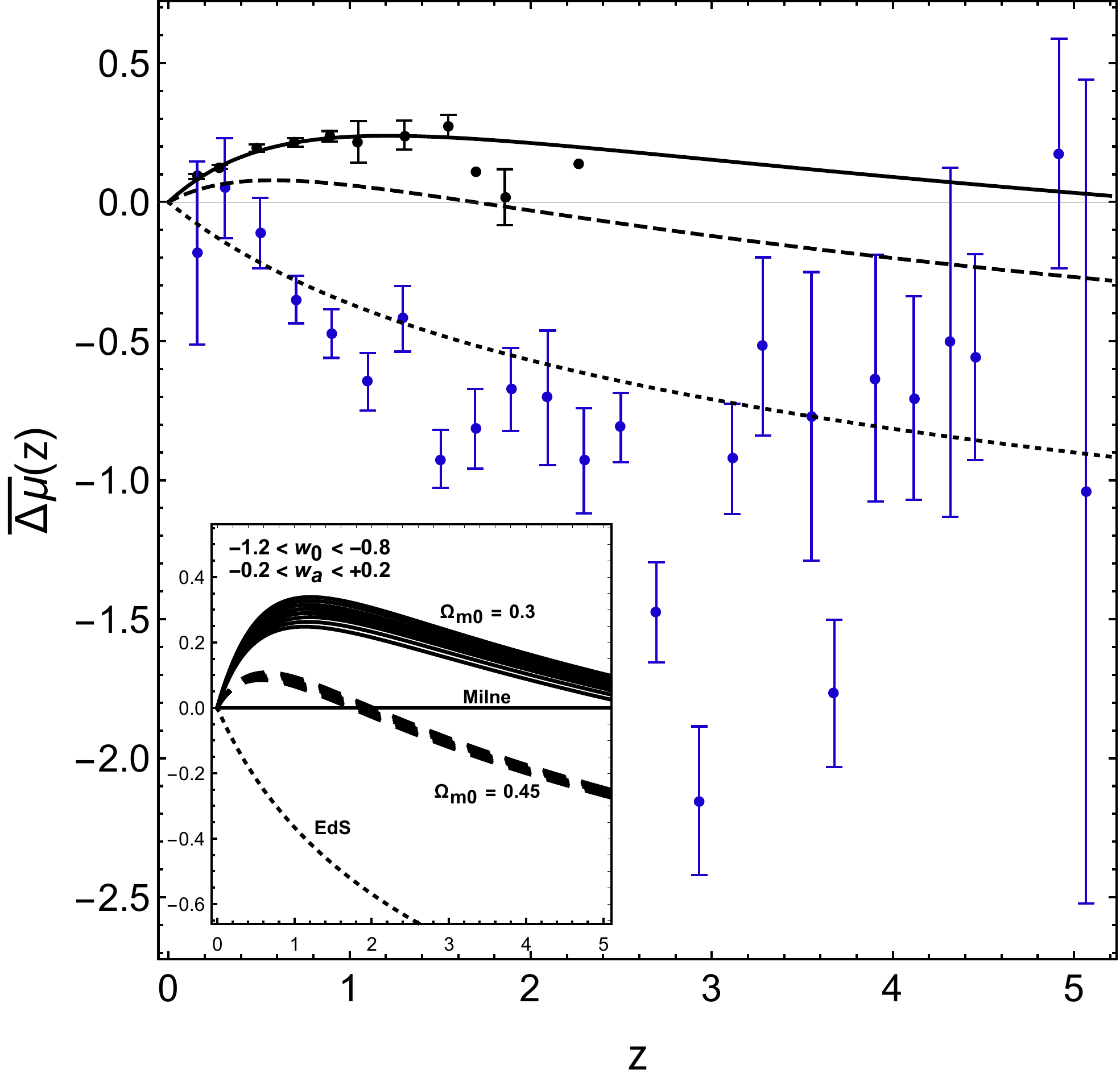}
\caption{Magnitude $\Delta \mu$ averaged over redshift bins of width $0.2$ for SN PANTHEON (black) sample and QSOsHD (blue). The inset shows the theoretical evolution of the estimator $\Delta \mu$ as a function of the redshift. Solid lines adopt $\Omega_{m0}=0.3$ and remain at positive $\Delta \mu$ values for the interval $0< z < 5$. Cosmologies adopting $\Omega_{m0}=0.45$ are plotted in the dashed lines. The dotted line represents the Einstein de-Sitter cosmology $\Omega_{m0}=1$. Both set of curves plotted in the inset assume parameters values in the range $-1.2 < w_0 < -0.8$ and $-0.2 < w_a < +0.2$. The solid and dashed lines shown with the binned data assumed the $\Lambda$CDM model.  }
\label{Fig2}
\end{figure}

We show in Fig. \ref{Fig2} the $\Delta \mu$ values for binned (bin width 0.2) data. Black (Blue) data points stands for the binned Pantheon SNHD (QSOsHD). We have performed this with different bin configuration and the general aspect of the binned data remains the same. The two black points in the SNHD with no error bars represent bins with a single SN. We also plot the expected (theoretical) evolution of the estimator (8) which can be computed by replacing $\mu_{obs}$ by $\mu_{th}$ where the latter is calculated with (\ref{mu}) for a given cosmological model.
 Solid (Dashed) lines assume $\Omega_{m0}=0.3$ $(\Omega_{m0}=0.45)$. Both curves assume a cosmological constant behavior for dark energy. The Milne universe is represented by the horizontal line at $\Delta \mu =0$. The Einstein-de Sitter universe $\Omega_{m0}=1$ corresponds to the dotted line fully in the region $\Delta \mu <0$.
The inset in Fig. \ref{Fig2} shows the expected $\Delta \mu$ for different cosmologies. It is assumed a dark energy equation of state parametrized by the CPL equation of state (EoS) parameter $w_{de}=p_{de}/\rho_{de}=w_0+w_1(1-a)$, where $p_{de}$ $(\rho_{de})$ is the dark energy pressure (energy density). The quantity $\Delta \mu$ is plotted (in the inset) for various parameters values in the range $-1.2 < w_0 < -0.8$ and $-0.2 < w_1 < +0.2$. Of course the cosmological constant behavior is within this parameter range at $w_0=-1$ and $w_a=0$. For all possible EoS parameter value in the adopted range $\Delta \mu$ remains positive in the interval $0<z<5.1$ for $\Omega_{m0}=0.3$. Therefore one can not expect $\Delta \mu < 0$ in an accelerated cosmology with $\Omega_{m0}=0.3$ when data up to $z\sim 5$ is used. It is therefore worth noting that the results of Table \ref{TableII} demonstrate that the QSOsHD is inconsistent with the standard cosmology $\Omega_{m0}\sim 0.3$ value.

The dashed lines in Fig. \ref{Fig2} show however that $\Delta\mu < 0$ values are expected in a $\Omega_{m0}\sim 0.45$ universe if data in the range $2<z<5$ is available. A $\Lambda$CDM universo with $\Omega_{m0}=0.45$ is indeed accelerated today. In this case the transition redshift from the decelerated phase to the accelerated one is $z^{0.45}_{tr}=0.35$. This fact can therefore yield to the conclusion that there is a hint for $\Omega_{m0} \gtrsim 0.4$ in the QSOsHD as already mentioned in Ref. \cite{Risaliti:2018reu}. However, $\Delta \mu$ has only a weak dependence on the dark energy equation of state parameters. The estimator is indeed more sensitive to $\Omega_{m0}$ than $w_{de}$ values.

\section{Final Remarks}

The Hubble diagram of high redshift objects is a promising tool to increase our understanding the universe's evolution. Risaliti \& Lusso \cite{Risaliti:2018reu} reported recently a compilation of 1598 objects filling the Hubble diagram in the redshift range $0.5<z<5.1$.
We have analysed in this work this sample under the perspective of a model-independent estimator for the background accelerated expansion according the description done in section \ref{sect2}. This estimator provides whether or not there is positive evidence for acceleration in Hubble diagrams data sets. The main result of this work is to show that one can not infer from the Hubble diagram of Quasars that the universe experienced an accelerated phase along its evolution. 

We found that the estimator is weakly dependent on the dark energy equation of state. It is therefore not possible to state a tension with standard cosmology as claimed in Ref. \cite{Risaliti:2018reu}. There is indeed a hint for higher $\Omega_{m0}$ values in agreement with obtained in Ref. \cite{Risaliti:2018reu}. But this does not indicate a strong indication for a cosmological tension relating the $\Omega_{m0}$ parameter since the QSOsHD is apparently not self-consistent, via the $\Delta \mu$ test, when QSOs data in the interval $z<2$ is used. Again, even for $z<2$ data the QSOsHD does not provide evidence for acceleration. In this redshift interval, while one expects $\Delta \mu >0$ for a $\Omega_{m0}=0.45$ cosmology (see Fig. \ref{Fig2}) the inferred value from the sample yields to negative $\Delta \mu$ values (as in Table \ref{TableII}). This means that QSOs can not indicate accelerated expansion which is widely confirmed fact. The source of such discrepancy found in this work is related to the fact that the QSOsHD is still very scattered. Further refinements of the sample are necessary to build a trustful Hubble diagram.

{\bf Acknowledgments:} We thank E. Lusso for providing the QSOsHD data used in this work. We appreciated discussion with D. Schwarz. HV thanks CNPq and FAPES for partial financial support. SG thanks CAPES for financial support.


\begin{thebibliography}{99}

\bibitem{Aghanim:2018eyx} 
  N.~Aghanim {\it et al.} [Planck Collaboration],
  arXiv:1807.06209 [astro-ph.CO].

\bibitem{Phillips:1993ng} 
  M.~M.~Phillips,
  Astrophys.\ J.\  {\bf 413}, L105 (1993).
  doi:10.1086/186970

\bibitem{Demianski:2016zxi} 
  M.~Demianski, E.~Piedipalumbo, D.~Sawant and L.~Amati,
  Astron.\ Astrophys.\  {\bf 598}, A112 (2017)
  doi:10.1051/0004-6361/201628909
  [arXiv:1610.00854 [astro-ph.CO]].
  
 
\bibitem{Wei:2013xx} 
  J.~J.~Wei, X.~F.~Wu and F.~Melia,
  Astrophys.\ J.\  {\bf 772}, 43 (2013)
  doi:10.1088/0004-637X/772/1/43
  [arXiv:1301.0894 [astro-ph.HE]].
  

\bibitem{Izzo:2009bw} 
  L.~Izzo, S.~Capozziello, G.~Covone and M.~Capaccioli,
  Astron.\ Astrophys.\  {\bf 508}, 63 (2009)
  doi:10.1051/0004-6361/200912769
  [arXiv:0910.1678 [astro-ph.CO]].
  
\bibitem{Cardone:2009mr} 
  V.~F.~Cardone, S.~Capozziello and M.~G.~Dainotti,
  Mon.\ Not.\ Roy.\ Astron.\ Soc.\  {\bf 400}, no. 2, 775 (2009)
  doi:10.1111/j.1365-2966.2009.15456.x
  [arXiv:0901.3194 [astro-ph.CO]].
  

\bibitem{Liang:2008kx} 
  N.~Liang, W.~K.~Xiao, Y.~Liu and S.~N.~Zhang,
  Astrophys.\ J.\  {\bf 685}, 354 (2008)
  doi:10.1086/590903
  [arXiv:0802.4262 [astro-ph]].
  
  
\bibitem{Schaefer:2006pa} 
  B.~E.~Schaefer,
  Astrophys.\ J.\  {\bf 660}, 16 (2007)
  doi:10.1086/511742
  [astro-ph/0612285].


\bibitem{Risaliti:2018reu} 
  G.~Risaliti and E.~Lusso,
  Nat.\ Astron.\ 
  doi:10.1038/s41550-018-0657-z
  [arXiv:1811.02590 [astro-ph.CO]].


\bibitem{Lusso:2017hgz} 
  E.~Lusso and G.~Risaliti,
  Astron.\ Astrophys.\  {\bf 602}, A79 (2017)
  doi:10.1051/0004-6361/201630079
  [arXiv:1703.05299 [astro-ph.HE]].

\bibitem{Risaliti:2016nqt} 
  G.~Risaliti and E.~Lusso,
  Astron.\ Nachr.\  {\bf 338}, no. 2/3, 329 (2017)
  doi:10.1002/asna.201713351
  [arXiv:1612.02838 [astro-ph.CO]].



\bibitem{Risaliti:2015zla} 
  G.~Risaliti and E.~Lusso,
  Astrophys.\ J.\  {\bf 815}, 33 (2015)
  doi:10.1088/0004-637X/815/1/33
  [arXiv:1505.07118 [astro-ph.CO]].



\bibitem{Avni}
Y.~Avni and H.Tananbaum
Astrophys. J.{\bf305}, 83-99 (1986).

\bibitem{Salvestrini:2019thn} 
  F.~Salvestrini, G.~Risaliti, S.~Bisogni, E.~Lusso and C.~Vignali,
  arXiv:1909.12309 [astro-ph.GA].

\bibitem{Lusso:2019akb} 
  E.~Lusso, E.~Piedipalumbo, G.~Risaliti, M.~Paolillo, S.~Bisogni, E.~Nardini and L.~Amati,
  Astron.\ Astrophys.\  {\bf 628}, L4 (2019)
  doi:10.1051/0004-6361/201936223
  [arXiv:1907.07692 [astro-ph.CO]].


\bibitem{Nielsen:2015pga} 
  J.~T.~Nielsen, A.~Guffanti and S.~Sarkar,
  Sci.\ Rep.\  {\bf 6}, 35596 (2016)
  doi:10.1038/srep35596
  [arXiv:1506.01354 [astro-ph.CO]].
  
\bibitem{Shariff:2015yoa} 
  H.~Shariff, X.~Jiao, R.~Trotta and D.~A.~van Dyk,
  Astrophys.\ J.\  {\bf 827}, no. 1, 1 (2016)
  doi:10.3847/0004-637X/827/1/1
  [arXiv:1510.05954 [astro-ph.CO]].


\bibitem{Seikel:2007pk} 
  M.~Seikel and D.~J.~Schwarz,
  JCAP {\bf 0802}, 007 (2008)
  doi:10.1088/1475-7516/2008/02/007
  [arXiv:0711.3180 [astro-ph]].

\bibitem{Seikel:2008ms} 
  M.~Seikel and D.~J.~Schwarz,
  JCAP {\bf 0902}, 024 (2009)
  doi:10.1088/1475-7516/2009/02/024
  [arXiv:0810.4484 [astro-ph]].


 
\bibitem{Scolnic:2017caz} 
  D.~M.~Scolnic {\it et al.},
  Astrophys.\ J.\  {\bf 859}, no. 2, 101 (2018)
  doi:10.3847/1538-4357/aab9bb
  [arXiv:1710.00845 [astro-ph.CO]].

\bibitem{Velten:2017ire} 
  H.~Velten, S.~Gomes and V.~C.~Busti,
  Phys.\ Rev.\ D {\bf 97}, no. 8, 083516 (2018)
  doi:10.1103/PhysRevD.97.083516
  [arXiv:1801.00114 [astro-ph.CO]].




\bibitem{Santos:2007pp} 
  J.~Santos, J.~S.~Alcaniz, N.~Pires and M.~J.~Reboucas,
  Phys.\ Rev.\ D {\bf 75}, 083523 (2007)
  doi:10.1103/PhysRevD.75.083523
  [astro-ph/0702728].

\bibitem{Gong:2007zf} 
  Y.~Gong, A.~Wang, Q.~Wu and Y.~Z.~Zhang,
  JCAP {\bf 0708}, 018 (2007)
  doi:10.1088/1475-7516/2007/08/018
  [astro-ph/0703583 [ASTRO-PH]].
  
  
\end{thebibliography}
\end{document}